\newcommand{\CLASSINPUTtoptextmargin}{19mm}%
\newcommand{\CLASSINPUTbottomtextmargin}{43mm}%
\newcommand{\CLASSINPUTinnersidemargin}{12.9mm}%
\newcommand{\CLASSINPUToutersidemargin}{12.9mm}%
\documentclass[conference,10pt,a4paper]{IEEEtran}
\usepackage{amsmath}
\usepackage{times}
\usepackage{graphicx}
\usepackage{multirow}
\usepackage[none]{hyphenat}
\usepackage{float}
\usepackage{subfig}
\usepackage[nolist]{acronym}
\begin{acronym}[C-ITS]
\acro{C-ITS}{Cooperative Intelligent Transport Systems}
\end{acronym}
\begin{acronym}[CACC]
\acro{CACC}{Cooperative Adaptive Cruise Control}
\end{acronym}
\begin{acronym}[CCAM]
\acro{CCAM}{Cooperative, Connected and Automated Mobility}
\end{acronym}
\begin{acronym}[ETSI]
\acro{ETSI}{European Telecommunications Standards Institute}
\end{acronym}
\begin{acronym}[GNSS]
\acro{GNSS}{Global Navigation Satellite System}
\end{acronym}
\begin{acronym}[ISAC]
\acro{ISAC}{Integrated Sensing and Communication}
\end{acronym}
\begin{acronym}[JRC]
\acro{JRC}{Joint Radar and Communication}
\end{acronym}
\begin{acronym}[mmWave]
\acro{mmWave}{Millimiter Wave}
\end{acronym}
\begin{acronym}[OFDM]
\acro{OFDM}{Orthogonal Frequency Division Multiplexing}
\end{acronym}
\begin{acronym}[PCM]
\acro{PCM}{Platooning Control Message}
\end{acronym}
\begin{acronym}[RadCom]
\acro{RadCom}{Radar-based Communication}
\end{acronym}
\begin{acronym}[UAV]
\acro{UAV}{Unmanned Aerial Vehicle}
\end{acronym}
\begin{acronym}[V2I]
\acro{V2I}{Vehicle-to-Infrastructure}
\end{acronym}
\begin{acronym}[V2V]
\acro{V2V}{Vehicle-to-Vehicle}
\end{acronym}
\begin{acronym}[V2X]
\acro{V2X}{Vehicle-to-Anything}
\end{acronym}
\begin{acronym}[VANET]
\acro{VANET}{Vehicular Ad hoc Network}
\end{acronym}
\usepackage{xcolor}
\usepackage{url}
\usepackage{tabularx}
\makeatletter

\def\@maketitle{\newpage
\bgroup\par\addvspace{0.5\baselineskip}\centering%
\ifCLASSOPTIONtechnote
   {\bfseries\large\@IEEEcompsoconly{\sffamily}\@title\par}\vskip 1.3em{\lineskip .5em\@IEEEcompsoconly{\sffamily}\@author
   \@IEEEspecialpapernotice\par{\@IEEEcompsoconly{\vskip 1.5em\relax
   \@IEEEtitleabstractindextextbox{\@IEEEtitleabstractindextext}\par
   \hfill\@IEEEcompsocdiamondline\hfill\hbox{}\par}}}\relax
\else
   \vskip0.2em{\EuMWtitlesize\ifCLASSOPTIONtransmag\bfseries\LARGE\fi\@IEEEcompsoconly{\sffamily}\@IEEEcompsocconfonly{\normalfont\normalsize\vskip 2\@IEEEnormalsizeunitybaselineskip
   \bfseries\Large}\@title\par}\vskip1.0em\par
   \ifCLASSOPTIONconference%
      {\@IEEEspecialpapernotice\mbox{}\vskip\@IEEEauthorblockconfadjspace%
       \mbox{}\hfill\begin{@IEEEauthorhalign}\@author\end{@IEEEauthorhalign}\hfill\mbox{}\par}\relax
   \else
      \ifCLASSOPTIONpeerreviewca
         {\@IEEEcompsoconly{\sffamily}\@IEEEspecialpapernotice\mbox{}\vskip\@IEEEauthorblockconfadjspace%
          \mbox{}\hfill\begin{@IEEEauthorhalign}\@author\end{@IEEEauthorhalign}\hfill\mbox{}\par
          {\@IEEEcompsoconly{\vskip 1.5em\relax
           \@IEEEtitleabstractindextextbox{\@IEEEtitleabstractindextext}\par\hfill
           \@IEEEcompsocdiamondline\hfill\hbox{}\par}}}\relax
      \else
         \ifCLASSOPTIONtransmag
           {\@IEEEspecialpapernotice\mbox{}\vskip\@IEEEauthorblockconfadjspace%
            \mbox{}\hfill\begin{@IEEEauthorhalign}\@author\end{@IEEEauthorhalign}\hfill\mbox{}\par
           {\vspace{0.5\baselineskip}\relax\@IEEEtitleabstractindextextbox{\@IEEEtitleabstractindextext}\vspace{-1\baselineskip}\par}}\relax
         \else
           {\lineskip.5em\@IEEEcompsoconly{\sffamily}\sublargesize\@author\@IEEEspecialpapernotice\par
           {\@IEEEcompsoconly{\vskip 1.5em\relax
            \@IEEEtitleabstractindextextbox{\@IEEEtitleabstractindextext}\par\hfill
            \@IEEEcompsocdiamondline\hfill\hbox{}\par}}}\relax
         \fi
      \fi
   \fi
\fi\par\addvspace{0.0\baselineskip}\egroup}

\def\EuMWtitlesize{\@setfontsize{\EuMWtitlesize}{24}{24pt}}
\def\EuMWauthorsize{\@setfontsize{\EuMWauthorsize}{11}{11pt}}
\def\EuMWaffilsize{\@setfontsize{\EuMWaffilsize}{10}{10pt}}
\def\EuMWcaptionsize{\@setfontsize{\EuMWcaptionsize}{9}{10pt}}
\def\EuMWbibsize{\@setfontsize{\EuMWbibsize}{8}{10pt}}

\def\@IEEEauthorblockNstyle{\EuMWauthorsize\@IEEEcompsocnotconfonly{\sffamily}\@IEEEcompsocconfonly{\large}}
\def\@IEEEauthorblockAstyle{\EuMWaffilsize\@IEEEcompsocnotconfonly{\sffamily}\@IEEEcompsocconfonly{\itshape}\@IEEEcompsocconfonly{\large}}
\def\@IEEEauthordefaulttextstyle{\EuMWauthorsize\@IEEEcompsocnotconfonly{\sffamily}\sublargesize}

\def\thebibliography#1{\section*{\refname}%
    \addcontentsline{toc}{section}{\refname}%
    \EuMWbibsize\@IEEEcompsocconfonly{\small}\vskip 0.3\baselineskip plus 0.1\baselineskip minus 0.1\baselineskip
    \list{\@biblabel{\@arabic\c@enumiv}}%
    {\settowidth\labelwidth{\@biblabel{#1}}%
    \leftmargin\labelwidth
    \advance\leftmargin\labelsep\relax
    \itemsep \IEEEbibitemsep\relax
    \usecounter{enumiv}%
    \let\p@enumiv\@empty
    \renewcommand\theenumiv{\@arabic\c@enumiv}}%
    \let\@IEEElatexbibitem\bibitem%
    \def\bibitem{\@IEEEbibitemprefix\@IEEElatexbibitem}%
\def\newblock{\hskip .11em plus .33em minus .07em}%
\ifCLASSOPTIONtechnote\sloppy\clubpenalty4000\widowpenalty4000\interlinepenalty100%
\else\sloppy\clubpenalty4000\widowpenalty4000\interlinepenalty500\fi%
    \sfcode`\.=1000\relax}

\long\def\@makecaption#1#2{%
\ifx\@captype\@IEEEtablestring%
\par\@IEEEtabletopskipstrut
\else
\@IEEEfigurecaptionsepspace
\fi
\setbox\@tempboxa\hbox{\normalfont\footnotesize {#1.}\nobreakspace\nobreakspace #2}%
\ifdim \wd\@tempboxa >\hsize%
\setbox\@tempboxa\hbox{\normalfont\footnotesize {#1.}\nobreakspace\nobreakspace}%
\parbox[t]{\hsize}{\normalfont\footnotesize\noindent\unhbox\@tempboxa#2}%
\else
\ifCLASSOPTIONconference \hbox to\hsize{\normalfont\footnotesize\hfil\box\@tempboxa\hfil}%
\else \hbox to\hsize{\normalfont\footnotesize\box\@tempboxa\hfil}%
\fi\fi
\ifx\@captype\@IEEEtablestring%
\@IEEEtablecaptionsepspace
\else
\fi}

\newlength\tablecaptiontotableskip
\newlength\figuretocaptionskip
\def\@IEEEfigurecaptionsepspace{\vskip\figuretocaptionskip\relax}%
\def\@IEEEtablecaptionsepspace{\vskip\tablecaptiontotableskip\relax}%

\def\abstract{\normalfont%
\@IEEEabskeysecsize\bfseries\textit{\abstractname}\,\bfseries\textit{---}\,%
\@IEEEgobbleleadPARNLSP}%

\def\IEEEkeywords{\normalfont%
\@IEEEabskeysecsize\bfseries\textit{\IEEEkeywordsname}\,\bfseries\textit{---}\,%
\@IEEEgobbleleadPARNLSP}%
\def\endIEEEkeywords{\relax\vspace{0.67ex}%
\par\if@twocolumn\else\endquotation\fi%
\normalsize\normalfont}%

\DeclareRobustCommand*{\EuMWauthorrefmark}[1]{\raisebox{0pt}[0pt][0pt]{\textsuperscript{\footnotesize{#1}}}}%
\def\@IEEEauthorblockNtopspace{0ex}
\def\@IEEEauthorblockAtopspace{1mm}
\def\tablename{Table}
\def\thetable{\arabic{table}}
\def\IEEEkeywordsname{Keywords}
\def\subsubsection{\@startsection{subsubsection}{3}{\z@}{1.5ex plus 1.5ex minus 0.5ex}%
{0.7ex plus .5ex minus 0ex}{\normalfont\normalsize\itshape}}%
\newlength{\CPheadmatchindent}%
\def\@seccntformat#1{\hbox to\CPheadmatchindent{\csname the#1dis\endcsname}\hskip 0.1em \relax}
\IEEEilabelindentA \parindent
\IEEEilabelindent \IEEEilabelindentA
\IEEEelabelindent \parindent
\IEEEdlabelindent \parindent
\IEEElabelindent \parindent
\makeatother

\begin{document}
\raggedbottom
%
%
%
\title{On RadCom channel capacity for V2V applications}
%
%
\author{%
\IEEEauthorblockN{%
Elena Haller\EuMWauthorrefmark{\#1}, 
Oscar Amador\EuMWauthorrefmark{\#2}, 
Emil Nilsson\EuMWauthorrefmark{\#3}
}
\IEEEauthorblockA{%
\EuMWauthorrefmark{\#}School of Information Technology, Halmstd University, Sweden\\
\{\EuMWauthorrefmark{1}elena.haller,\EuMWauthorrefmark{2}oscar.molina, \EuMWauthorrefmark{3}emil.nilsson\}@hh.se\\
}
}
%
\maketitle
%
%
\begin{abstract}
The use of \ac{mmWave} for communication and sensing purposes is one of the functions powered by Next Generation \ac{V2X} networks. The arrival of IEEE~802.11bd, which is able to operate in the 60\,GHz band, opens the doors of \ac{ISAC} to vehicular networks. Similarly, \ac{RadCom} proposes the use of the radar spectrum for communication puproses. In this paper, we perform an analysis of the channel capacity for different configurations of RadCom, showing its potential to offload the V2X spectrum for bumper-to-bumper \ac{V2X} applications. We finalize with a discussion on the potential for ISAC from both the 802.11bd and RadCom approaches.

\end{abstract}
\begin{IEEEkeywords}
Channel capacity, Integrated Sensing and Communication, MIMO Radar, Vehicle-to-vehicle communication, Vehicular ad hoc networks.
\end{IEEEkeywords}
%
%

\section{Introduction}

\ac{CCAM} is the final stage of the road towards \textit{future mobility}, where road users exchange information and cooperate with each other to use the road safely and efficiently. This cooperation enables also sustainable mobility in terms of energy and human resources, and as such is recognized by the United Nations' Social Development Goals~\cite{Agenda2030}. Therefore, global initiatives such as Vision Zero~\cite{ecVisionZero} present a road map towards future mobility, when \ac{CCAM} is enabled in all roads and at all times.

Safety and efficiency is empowered by \ac{C-ITS}, that rely on \acp{VANET} to allow road users communicate with each other using \ac{V2X} communications. These networks have been operating in the 5.9\,GHz spectrum that is globally reserved for vehicular safety applications, and standardization organisms such as \ac{ETSI} have been developing protocols and frameworks such as the ETSI ITS framework.

The radio spectrum for \acp{VANET} can be used by different medium access technologies, and so far WiFi (i.e., IEEE802.11p known in ETSI as ITS-G5) and cellular (i.e., LTE and 5G) have been deployed at different levels and intensities~\cite{C-ITS2016,C-ROADS}. However, ETSI ITS considers the possibility of also using other access technologies and account for advances in technological capabilities. This is reflected in protocols that are media independent~\cite{etsiNewGeoNetworking}, the need for harmonization and coexistence between access technologies~\cite{etsiCoex}, and the prevision that future, advanced services will require extended capabilities~\cite{etsiMCO}.

One of these advanced services is Maneuver Coordination, and one of its subsets, platooning~\cite{etsiMCO}. The \ac{VANET} recognizes the need for more radio resources to enable future and current services to access the medium. Multi-channel operation is the first step to enable second generation services, and newer versions of WiFi and Cellular \ac{V2X} open a new door, the use of the \ac{mmWave} spectrum. IEEE~802.11bd offers the possibility of operating in the 60\,GHz spectrum to communicate, which opens the door for also adding positioning and sensing capabilities as is the case for other WiFI versions~\cite{Pika11az}. Furthermore, 6G also considers the use of \ac{mmWave} to perform \ac{ISAC}~\cite{ISaC6G}. 

Similarly, the use of technologies normally used for sensing, e.g., radar, to carry out communication tasks looks at \ac{ISAC} from the other end. \ac{RadCom} offers the possibility to turn sensors already existing in vehicles into network interfaces. Our previous work~\cite{haller2024offloading} explores the use of RadCom as an alternative to offload intra-platoon messages from the 5.9\,GHz channel to bumper-to-bumper communications enabled by RadCom. We showed that offloading \acp{PCM} to \ac{V2V} enables shorter inter-vehicle distances in a platoon. However, there remains an open question on the comparison between RadCom and other mmWave technologies such as 802.11bd in the 60\,GHz spectrum.

In this paper, we present an analysis of the medium capacity for two fronts of \ac{ISAC}: 802.11bd and RadCom. The contributions of this paper are: 
\begin{enumerate}
    \item An analytical study of the throughput for both technologies in bumper-to-bumper scenarios in a platoon.
    \item An exploration of different modulation schemes for RadCom.
    \item A comparison of these schemes in RadCom to the established mechanisms for 802.11bd.
\end{enumerate}

The rest of the paper is organized as follows: background and related work is presented in Section~\ref{sec:related_work}, channel capacity estimations are presented in Section~\ref{Sec:Channel_capacity}, numerical results for channel capacities of 802.11bd and RadCom are shown in Section~\ref{sec:numerical_results}, and a discussion for the capability of ISAC to support new V2V applications is presented in Section~\ref{sec:discussion}. Finally, conclusions and future work are in Section~\ref{sec:conclusion}.

\section{Related Work}
\label{sec:related_work}

In this section, we provide an overview of the state of the art on the use of \ac{mmWave} both separately for communications and for sensing and as a combination in \ac{ISAC}. We start by presenting the key points for each concept, and then we perform an analysis of relevant works in the literature.

\subsection{Integrated Sensing and Communications}
\label{subsec:ISAC}

Resource availability, specifically radio spectrum, is the main motivator for \ac{ISAC}. Spectrum that is allocated, e.g., for radar use, is a prime candidate to double as communication channels in an approach that is called \ac{JRC}~\cite{jrc2020}. This approach has also the benefit of turning mutual interference into a collaborative system where former competitors now share resources~\cite{isacsurvey}.

\begin{figure}
    \centering
    \includegraphics[width=0.8\linewidth]{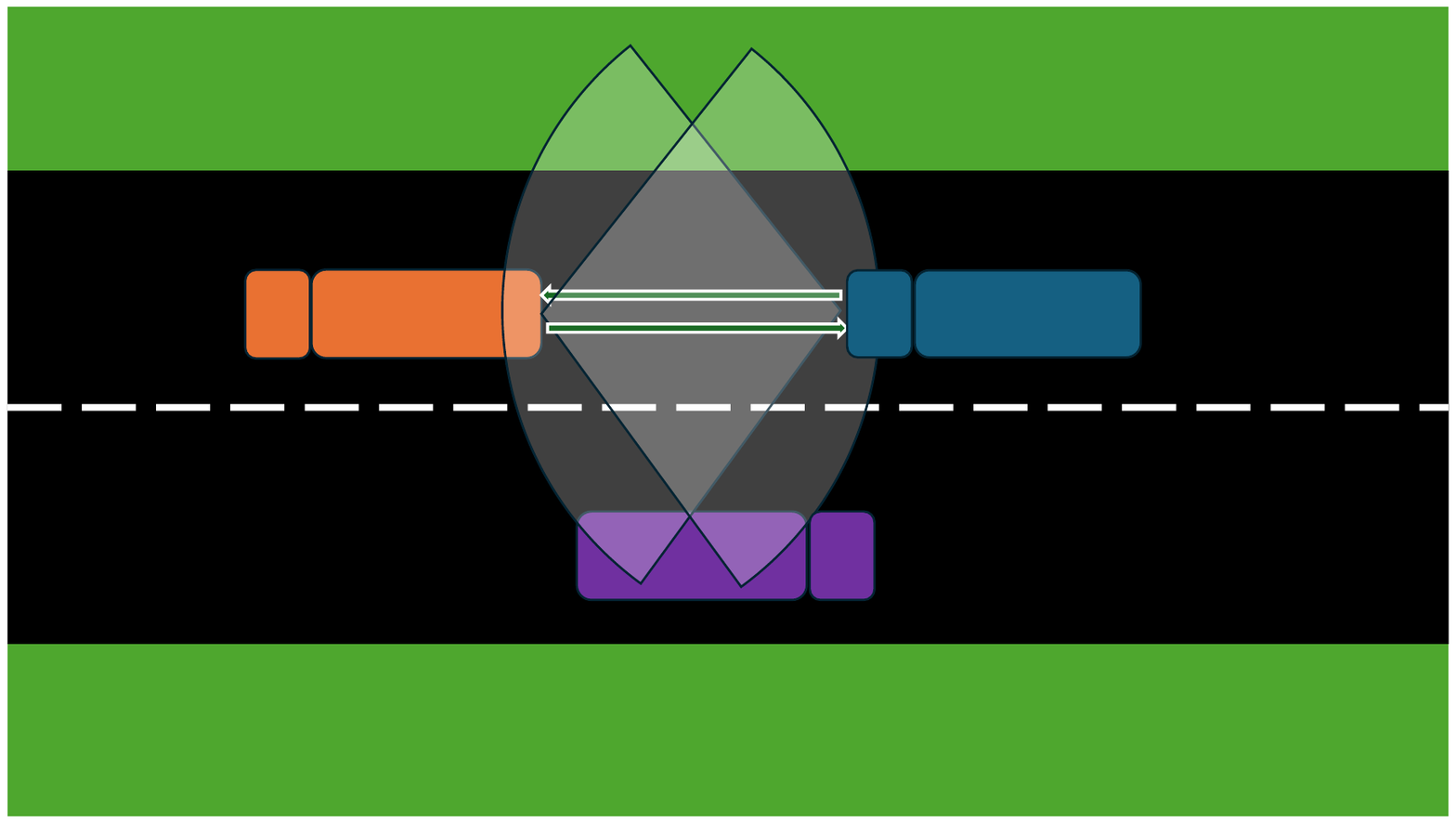}
    \caption{Integrated Sensing and Communications between two heavy-duty vehicles.}
    \label{fig:isac_platoon}
\end{figure}

Fig.~\ref{fig:isac_platoon} shows a scenario for \ac{ISAC}. The trucks drving towards the left have hardware (e.g., front-facing and back-facing radars, or \ac{mmWave} antennae) than enables them to 1) sense their relative distance (white shadow), and also detect the purple truck; and 2) exchange information through a communication link (arrows between the vehicles). These functions can occur simultaneously or in an alternating fashion.

There are two identifiable fronts for \ac{ISAC}. While \ac{JRC} has the goal of \textit{piggybacking} communications on a signal meant for sensing, there is also an approach where network signals can double as \textit{radars} or sensors to power computer vision. One example for general use cases is the proposed ability for 6G cellular networks to use \ac{mmWave} to offer sensing capabilities, e.g., at base stations~\cite{ISaC6G}. 

The literature explores \ac{ISAC} in the \ac{mmWave} range in dynamic scenarios. The work in~\cite{isacuav} explores the use of \ac{ISAC} for both locating and controlling \acp{UAV}. The work analyzes whether terahertz signals cover the requirements for sensing and communication. This work is relevant since it addresses a system level perspective, where kinematics affect network performance, and at the same time can be controlled by sensing. \ac{ISAC} in cooperative networks is studied in~\cite{coopISAC}. The authors present an abstract scenario where a source node transmits information using an \ac{ISAC} signal to a destination with the help of a relay node. Their results show that there shall be a balance between sensing and communication so that functions do not disrupt each other. These works are relevant for the specific case we study in this paper, where \acp{PCM} are offloaded from the main channel in the 5.9\,GHz spectrum into \ac{RadCom}. While the minimum requirements for ensuring safety in a platoon can be ensured by exchanging \ac{V2V} messages between consecutive vehicles~\cite{Galina-ethics}, a remaining question is whether more platooning services can be offloaded to bumper-to-bumper, multi-hop communications between platoon members. As a firs step, we study the potential performance of two types of ISAC-capable access technologies: vehicular WiFi (i.e., IEEE~802.11bd) and RadCom.

\subsection{IEEE802.11bd on the mmWave band}
\label{subsec:802.11bd}

\begin{table}[http!]
    \centering
    \caption{Specifications for 802.11p and 802.11bd}
    \label{tab:specs_11}
    \begin{tabular}{c|c|c}
     \textbf{Feature} & \textbf{802.11p} & \textbf{802.11bd}  \\
     \hline
     Frequency bands & 5.9\,GHz & \{5.9, 60\}\,GHz \\
     System bandwidth (at 5.9\,GHz) & 10\,MHz & \{10, 20\}\,MHz \\
     Data subcarriers (at 5.9\,GHz) & 48 & 48 \& 52 \\
     MIMO & N/A & $2\times2$ MIMO \\
     Data rates (in 10\,MHz) & 3 to 27\,Mbps\dag & Up to 39\,Mbps \\
     Relative veh. speed & 252~km/h & 500~km/h \\
     mmWave (60\,GHz) & Not supported & Supported \\
     Localization & N/A & Supported \\
    \end{tabular}\\
    \raggedright \dag Only 6\,Mpbs is typically used.
\end{table}

For over a decade, IEEE~802.11p (adopted by \ac{ETSI} as ETSI~ITS~G5) has been the standard WiFi-based access technology for \acp{VANET}~\cite{etsig5}. Following advances in other WiFi technologies, an evolution for WiFi-based \ac{V2X} known as IEEE~802.11bd was launched in early 2023~\cite{80211bd2023}. Table~\ref{tab:specs_11} shows an overview of the differences between the two versions of vehicular 802.11. The two core feature changes in 802.11bd are 1) the ability to use twice as much bandwidth as 802.11p, and 2) the use of the unlicensed 60\,GHz spectrum, which brings the possibility of using mmWave-related features such as positioning. However, these two features open two main questions: i) whether the need for backwards compatibility with 802.11p will render the 20\,MHz bandwidth unusable, and ii) if the mmWave features (which are inherited from 802.11ay, designed for indoor use) work in highly-dynamic, outdoors vehicular scenarios~\cite{11bdFrom11ay}.

The work in~\cite{SolsBDfromAD} explores these challenges, namely, overhead stemming from beam training, loss of line-of-sight, misalignment, range limitations, and delay. The authors identify problems in vehicular scenarios where mmWave communications are used in generic \ac{V2X} use cases. They propose the use of relay vehicles to enable \ac{V2V} and \ac{V2I} communications when the ego vehicle experiences communication problems, e.g., beam misalignment or loss of line-of-sight. Finally, they propose the use of offline information (e.g., high-definition maps) to speed up beamforming. 

The ability of WiFi in the 60\,GHz band to provide precise positioning information is explored in~\cite{Pika11az}. There, authors experiment with 802.11az and obtain accuracy at the centimeter level when: 1) nodes have time synchronization, 2) they are in line-of-sight, and 3) they are able to identify the \textit{first path}. Furthermore, even if the scope of the work is indoor scenarios, the distances they explore are relevant for our use cases, where two vehicles might be between a close distance to each other, and multi-path phenomena can affect performance.

\subsection{Radar-based Communications}
\label{subsec:related_radcom}

\ac{RadCom}, or \ac{JRC}, is the approach to \ac{ISAC} that originates on radar signals. The pervasiveness of radar devices in vehicles with or without any automated driving capabilities allows for a higher chance for equipping vehicles with mmWave communications from the get-go.

The use of \ac{RadCom} for general applications has been widely explored in the literature. The work in~\cite{handRadCom} assesses the possibility of simultaneous sensing and communication using RadCom to detect hand movements and share the information with near devices. They achieve data rates in the range of gigabits but at distances below 5~cm. However, automotive applications require longer distances and more dynamic environments, where vehicles/nodes are also in movement.

In a similar fashion as in WiFi in the 60\,GHz spectrum, time synchronization is also a concern in RadCom. Authors in~\cite{radcomSync} analyze a setup that is viable in vehicular scenarios: obtaining timing information from a \ac{GNSS}. Results show that, even with references from \ac{GNSS}, the timing services in nodes drift. Furthermore, from our system point of view, this issue is potentially more present in the particular scenarios where ISAC is needed to improve tracking and positioning of neighbors.


\subsection{Related Work on Channel Capacity}
\label{subsec:related_capacity}

Studies on channel capacity for mmWave deployments exist in the literature. The work in~\cite{mmWaveCap} explores channel capacity in mmWave at two bands (28 and 73\,GHz). Using Monte Carlo simulations as a function of signal to noise ratio, they obtain capacities in the range of hundreds of megabits. The study in~\cite{mmWaveUrban} presents an analysis of channel capacity for mmWave with urban micro and macro channel models. A numerical evaluation of four bands (28, 38, 60, and 73\,GHz) shows that mmWave has a channel capacity in the order of gigabits although with a limited range due to path loss. Furthermore, the authors in~\cite{mmWaveMIMO} perform a study on the effect of antennae array sparsity on signal degradation and communication capacity, and how the addition of irregular antennae arrays helps improve performance. On a similar note, the work in~\cite{mmWaveMassiveRelay} explores different deployments to counteract the effect of path loss on system throughput by using relay nodes to help signals reach their destination via artificial propagation paths. These works, explore the effect of distance on mmWave channel capacity, however, they do not account for the dynamism of vehicular networks, where sources and destinations are in constant movement. The work in~\cite{mmWaveV2I} does explore channel capacity for mmWave in vehicular scenarios, but only for \ac{V2I} scenarios, where the infrastructure nodes are static. Furthermore, this work is focused in the sub-6\,GHz band. Finally, these works exploring the capacity of the mmWave spectrum do not account for \ac{ISAC} scenarios, although their results are a beacon on the potential for the communication capacity of the mmWave band, where \ac{ISAC} resides.

Similarly, channel capacity and resource allocation has been studied for \ac{RadCom}. The work in~\cite{radcomPartition} explores the assignment of antenna elements in an array to sensing or communication functions. Their results show that assigning elements randomly and dynamically allow for large channel capacities (in the order of Mbits/Hz) while keeping a high sensing resolution (similar to that of a full antennae array). However, their study covers only the 9\,GHz band. In a similar fashion, the authors in~\cite{radcomCapacity5-8} assess sensing and communication capabilities for RadCom in the 5.8\,GHz band. They propose an architecture that communicates with multiple nodes and uses the interference between them to perform environmental sensing. Their results show a trade-off between sensing and communication. Finally, another work that explores the trade-off between sensing and communication in RadCom is presented in~\cite{radcomCapacityEM}. Here, authors perform a numerical analysis of RadCom in the 77\,GHz band and find a Pareto point for the dual-function system. In this work, we assess channel capacity for a RadCom system focusing solely on the communication functionality of \ac{ISAC}, e.g., on the effect of bandwidth and number of sub-carriers, and not on modulation schemes or channel models. 

\section{Channel capacity}
\label{Sec:Channel_capacity}

\begin{table*}[tbh!]
    \centering
    \caption{Parameters used for 802.11bd at 60\,GHz and RadCom}
    \label{tab:Matlab_values}
    \begin{tabularx}{\textwidth}{X|X|X|X|X}
     \textbf{Parameter}    & \textbf{Description}  & \textbf{RadCom} & \textbf{802.11bd} & \textbf{Unit}  \\
     \hline
        $G_T$  &Transmitter gain & 10 & 12.5 & dB \\ 
        $G_R$ & Receiver gain & 10 & 12.5 &dB \\
        $f_{min}$ & Lowest frequency & [\textbf{76}--81] & 60 &GHz \\
        $N_{CS}$ & Number of bins & \textbf{3276} [thousands] & 3276 & --- \\
        $BW$ & Bandwidth & $1.5\times10^8 - \mathbf{1\times10^9}$ & $1.280\times10^9$&Hz \\
        $c$ & Signal speed & $3\times 10^8$ & & m/s \\
        $F$ & Receiver's noise figure & $\mathbf{8}-10$ & 7.5 & dB \\
        $\delta$ & Duty cycle & 1.0 & 1.0 & --- 
    \end{tabularx}
\end{table*}

Channel capacity $C$ is a function of signal to noise ratio (SNR).
According to Shannon's equation, for the system operating at frequences $f\in[f_{min}, f_{max}]$ with bandwidth
$BW=f_{max}-f_{min}$, the channel capacity can be calculated as
\begin{equation}
\label{Cap_cont}    
C=\delta\int\limits_{f_{min}}^{f_{max}} \log_2\left(1+\frac{P_S}{P_N}\right)df.
\end{equation}
Here $\delta\in(0,1)$ is the duty cycle of the system and $P_S$ and $P_N$ represent signal and noise powers at the receiver correspondingly.

Both radar and wifi channels are in the range $0.5\times 10^7 \mathrm{Hz}<f<10^{12}\mathrm{Hz}$, where thermal noise dominates (see~\cite{oberg}, p.~24).  The noise power $P_N$ in this case can be estimated as
\[
P_N=kT_0FBW,
\]
where $k= 1.380649 \times 10^{-23}\mathrm{J/K}$ is Boltzmann's constant, $T_0$ is the temperature of the receiver (usually $290~K$) and $F$ is the receiver noise figure (see~\cite{IKnowTheREF}). 

The signal power $P_S$ at the receiver is calculated according to COST207 statistical model (see \cite{saleh}, \cite{IKnowTheREF}) with the mean path power gain of the first arrival $\alpha^2$ defined via Friis' equation as follows:
\begin{equation}\label{loss}
\alpha^2(f)= G_TG_R\frac{c^2}{4\pi d^2 f^2}.\end{equation}
Here
$G_T$ and $G_R$ are transmitter's and receiver's gains, $c$ is the signal speed (speed of light) and $d$ is the length of the channel.

For \ac{OFDM} modulation scheme with equally spaced $N_{CS}$ subcarriers of width $df=BW/N_{CS}$ the loss $\alpha^2(f)$ for the $n$-th bin, i.e. channel operating at $f\in[f_{n-1},f_{n})$, can be estimated as
\begin{equation}
\label{loss2}
\alpha^2(f_{n})<\alpha^2(f)<\alpha^2(f_{{n-1}}).
\end{equation}

Thus, to estimate the total capacity for \ac{OFDM} channel one can replace integration in \eqref{Cap_cont} with a finite sum over all $N_{SC}$ bins within the region $[f_{min}, f_{max}]$ and use loss estimates \eqref{loss}.

\section{Numerical Results}
\label{sec:numerical_results}

In order to compare capacities of WiFi and RadCom channels, we define numerical values for parameters in \eqref{Cap_cont} and \eqref{loss}. Table~\ref{tab:Matlab_values} summarizes the  values we use, which are in the range of those recommended by the International Telecommunications Union for automotive radar~\cite{ituRadarAut}.

In Section~\ref{Subsec:RadComConfig}, we explore various RadCom channel configurations. It results in a choice of optimal parameters (marked in bold in Table~\ref{tab:Matlab_values}).

The comparison between WiFi channel and the optimal RadCom setting is performed in Section \ref{Subsec:RadComWifi}. Since the duty cycle is not yet determined for Wifi sensing channels (further discussed in Sections \ref{Subsec:RadComWifi}, \ref{sec:discussion}), both WiFi and RadCom capacities are compared with no  $\delta$-scaling, i.e. for $\delta=1$.

\subsection{RadCom configuration analysis}
\label{Subsec:RadComConfig}

There are 5 parameters that allow several RadCom setups: $f_{min}$, $N_{CS}$, $BW$, $F$, $\delta$. 
\begin{itemize}
    \item[$f_{min}$:] 
    From  
\eqref{Cap_cont} and \eqref{loss} it follows that the particular choice for $f_{min}$ does not affect capacity calculations significantly. Indeed,
for limit values for $f_{min}$, $76$ and $81$ GHz along the distances $d=1, 50, 20$ (m) and equal values for the rest of parameters (marked in bold in Table~\ref{tab:Matlab_values}) the corresponding relative difference is less than 3 \% (see Table~\ref{tab:RadComfmin}).
\end{itemize}
\begin{table}[htp!]
    \centering
     \caption{RadCom channel capacity for $f_{min}=76 \mathrm{GHz}$ and $f_{min}=81 \mathrm{GHz}$}
    \label{tab:RadComfmin}
    \begin{tabular}{c|r|r|r}
    \textbf{d(m)}& \textbf{76 GHz: C\,(Gbps)}& \textbf{81 GHz: C\,(Gbps)}& $\boldsymbol{\Delta}$\textbf{C/C (\%)} \\
    \hline
      1   & 22.2078 &22.0251&  0.8226\\
       50  & 10.9208 & 10.7383 & 1.6718\\
       200 & 6.9320 &6.7509 &2.6123
    \end{tabular}
   
\end{table}

\begin{itemize}
\item[$N_{CS}$:] Due to the integral nature of capacity as system's characteristic, the number of bins doesn't change \eqref{Cap_cont} directly. However, low number of subcarries leads to larger estimation intervals in \eqref{loss2}, i.e. makes capacity estimates less precise. In Table~\ref{tab:RadComBins}, calculations for lower bound for $N_{CS}=10^2$ and $N_{CS}=10^4$ still show low differences (below $1 \%$). , see Table~\ref{tab:RadComBins}).
\end{itemize}

\begin{table}[htp!]
    \centering
     \caption{RadCom channel capacity for $N_{CS}=10^2$ and $N_{CS}=10^4 $}
    \label{tab:RadComBins}
    \begin{tabular}{c|r|r|r}
    \textbf{d(m)}& \textbf{10$^2$ bins:\,C (Gbps)}& \textbf{10$^4$ bins:\,C (Gbps)}& $\boldsymbol{\Delta}$\textbf{C/C (\%)} \\
    \hline
      1   & 22.2076 & 22.2078 &  -0.0008\\
       50  &  10.9207& 10.9208 & -0.0017\\
       200 & 6.9318& 6.9320 &-0.0026
    \end{tabular}
   
\end{table}

\begin{itemize}
\item[$BW$:] Channel capacity $C$ depends linearly on bandwidth $BW$ as the domain of integration in \eqref{Cap_cont}. It is the main parameter affecting $C$. Table~\ref{tab:RadComBW} presents the results for $BW=150$ MHz and  $BW=1$ GHz. The difference in capacity values lies around $80 \%$.
\end{itemize}

\begin{table}[tbh!]
    \centering
     \caption{RadCom channel capacity for $BW=150$\,MHz and $BW=1$\,GHz}
    \label{tab:RadComBW}
    \begin{tabular}{c|r|r|r}
    \textbf{d(m)} & \textbf{150\,MHz: C (bGps)}& \textbf{1\,GHz: C (Gbps)}& $\boldsymbol{\Delta}$\textbf{C/C (\%)} \\
    \hline
      1   & 3.7441& 22.2078& -83.1405\\
       50  &2.0509 & 10.9208&  -81.2195\\
       200 & 1.4512 & 6.9320& -79.0646
    \end{tabular}
\end{table}

\begin{itemize}
\item[$F$:] Receiver's noise figure $F$ which is reflected in the noise power $P_N$ and makes inverse ratio with capacity $C$. Table~\ref{tab:RadComF} demonstrates effects within $3\%$ -- $10\%$ for noise figures $F=8$ dB and $F=10$ dB.
\end{itemize}

\begin{table}[tbh!]
    \centering
     \caption{RadCom channel capacity for $F=8$ dB and $F=10$ dB}
    \label{tab:RadComF}
    \begin{tabular}{c|r|r|r}
    \textbf{d(m)}& \textbf{8\,dB: C (Gbps)}& \textbf{10\,dB: C (Gbps)}& $\boldsymbol{\Delta}$\textbf{C/C (\%)}  \\
    \hline
      1   &  22.2078 &   21.5434 & 3.0839\\
       50  &  10.9208 &   10.2569 & 6.4731\\
       200 &  6.9320 & 6.2745 & 10.4787
    \end{tabular}
\end{table}

\begin{itemize}
\item[$\delta$:] Duty cycle $\delta$ decreases capacity \eqref{Cap_cont} and defines the fraction that actually can be reached, i.g. for RadCom communication due to limitations by sensing, $\delta=0.1$ lowers the capacity value to $10\%$ of the one with continuous communication (when $\delta=1$). 
\end{itemize}

\subsection{RadCom and WiFi channel capacity}
\label{Subsec:RadComWifi}

\begin{figure}
    \centering
\includegraphics[width=\linewidth]{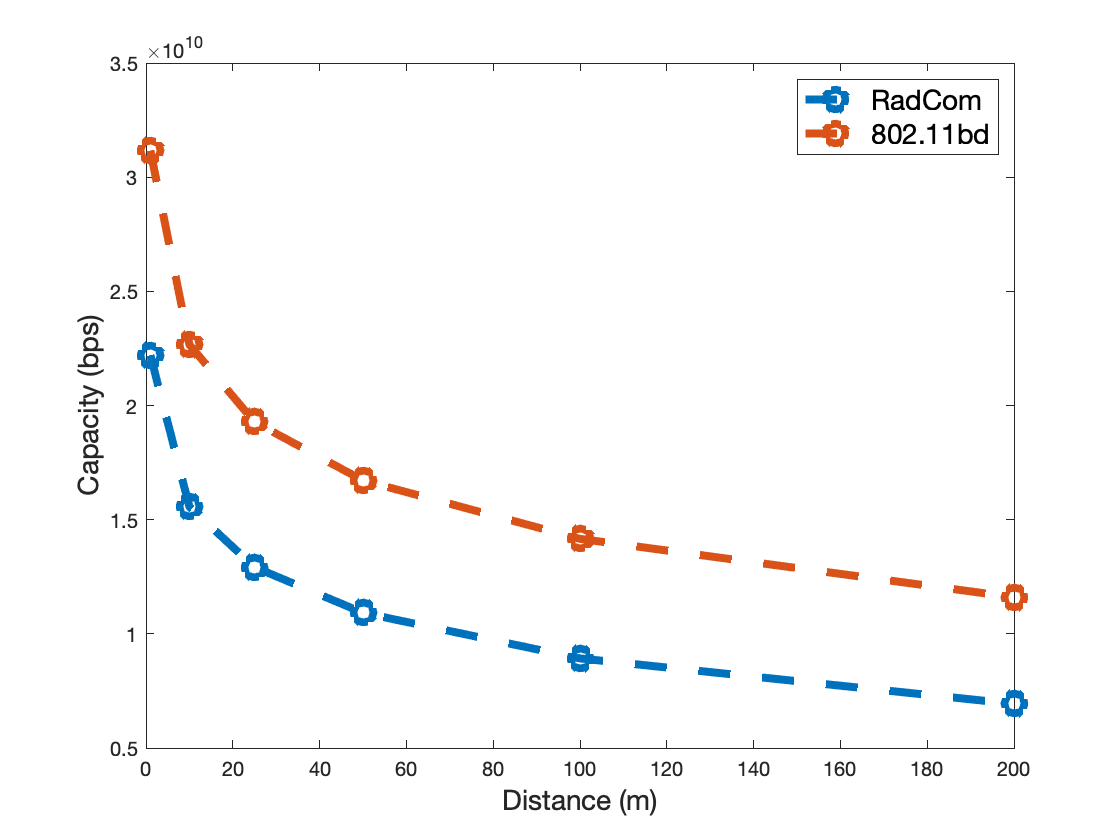}
    \caption{Channel capacity for different distances between nodes}
    \label{fig:enter-label}
\end{figure}

Fig.~\ref{fig:enter-label} shows the results for RadCom and WiFi communication channels with full duty cycles ($\delta=1$). RadCom parameters are optimized according to Section \ref{Subsec:RadComConfig} and denoted in bold in Table \ref{tab:Matlab_values}. Thus, the plot considers both channels to use 100\% of its bandwidth and time for communication functions. With this head-to-head comparison, both WiFi and RadCom ar able to keep a capacity over 10\,Gbps for up to 50\,m between nodes. 

When factoring in $\delta$ and overhead (e.g., from medium access techniques, signaling, collision avoidance), even if 90\% of that capacity is lost to $\delta$, and an extreme 50\% of the remainder is lost to overhead, the channel capacity stays still in the hundreds of Mbps for both WiFi and RadCom with inter-vehicle distances of up to 50\,m (e.g., feasible distances within a platoon), surpassing clearly the 6\,Mbps of 802.11p.



\section{Discussion}
\label{sec:discussion}

While our results show that 802.11bd in the 60\,GHz band outperforms RadCom even if we consider a scenario when the radar band is used completely for communications, the purpose of \ac{ISAC} is to combine sensing and networking functions. RadCom performs sensing natively, but the way this will be achieved in vehicular WiFi is still an open question. Versions of WiFi that offer positioning and sensing capabilities~\cite{Pika11az} usually perform these functions in three ways: 1) detecting other nodes using 802.11bd and measuring their distance using radio information (e.g., fine timing measurements), 2) measuring the interference between nodes to detect obstacles, and 3) a radar-like approach where a node sends out signals and listens to its reflections.

Approaches 1) and 2) require that other nodes also use 802.11bd in the 60\,GHz band for fine timing measurements to work and to detect propagation phenomena that can lead to identifying obstacles in the environment. This, in turn, requires time synchronization to perform, e.g., time-of-flight measurements. This can be challenging when considering time synchronization might be dependant on entities in different layers and different implementations of the standard and on the capabilities of the hardware where the framework runs.

For approach 3), time synchronization is not needed, since the node is listening to its own messages. However, it will require for it to stop communications in order to listen to its own transmission, thus, having a duty cycle like RadCom. Furthermore, the ability of WiFi to achieve the sensing performance of radar (e.g., resolution) is another possible trade-off. On this same note, even if 802.11bd-based ISAC might not replace radar as the go-to solution for vehicular sensing (i.e., if ISAC is used as an additional sensor), factors such as antenna placement have to be factored in to maximize performance in both networking and sensing functions.

The gap in capacity can be also be breached if RadCom leverages it advantages in a system-level. Works in the literature explore problems for mmWave regarding beam training, misalignment, and loss of line-of-sight, and these issues can be overcome if features like \ac{CACC}, which is supported by radar and networking. Being able to control the relative position between vehicles can allow for RadCom to use more ambitious modulation and schemes, and thus increase throughput.

Finally, there is not a visible reason why both 802.11bd and RadCom cannot coexist. In fact, RadCom opens the door for \textit{legacy} connected vehicles (e.g., running 802.11p) to enter the ISAC ecosystem. The calculated capacity for ISAC exceeds the one on the 5.9\,GHz band, and might enable services such as V2V see-through, exchanging of sensor information, maneuver coordination, and other applications that would require, e.g., high throughput or re-transmissions. Nevertheless, it is important to remember that even if a technology offers certain capabilities, the road to implementation might lead to more conservative approaches, as it happened with the ability of 802.11p to support several datarates and 6\,Mbps being the default, fixed datarate in its ETSI ITS-G5 implementation.


\section{Conclusion}
\label{sec:conclusion}

We present an analysis of channel capacity for two ISAC-capable, vehicular networking channels in the mmWave spectrum: 802.11bd in the 60\,GHz band and \ac{RadCom} at 76\,GHz. Our analysis shows that 802.11bd outperforms radar when using the channel exclusively for communications. We also explore the effect of different parameters (i.e., operational frequency, number of sub-carriers, bandwidth, and receiver noise figure) on the channel capacity of RadCom. The first takeaway of this exploration is that, even when factoring in a duty cycle of 0.1 for RadCom, channel capacity offers enough room to i) replace the the 5.9\,GHz band in bumper-to-bumper applications, and ii) offer a channel for new \ac{V2V} applications. Finally, we present a discussion on the ISAC abilities for 802.11bd and RadCom, as well as the advantages and disadvantages of each technology on a system level. Future work includes the exploration of different configurations for RadCom to find an optimal point where sensing and communication capabilities operate at a level which minimizes the trade-off between functionalities.


\section*{Acknowledgment}

This work was partially supported by SAFER in the project ``Human Factors, Risks and Optimal Performance in Cooperative, Connected and Automated Mobility'', the ELLIIT Strategic Research Network in the project ``6G wireless'' – sub-project ``vehicular communications'', and Vinnova grant 2021-02568.


\bibliographystyle{IEEEtran}

\bibliography{EuMW_Haller.bib}

\end{document}